\documentclass{svjour3}                     
\smartqed  
\usepackage{graphicx}
\usepackage{amsmath,amssymb,bm}
\begin{document}

\title{Chiral SU(3) dynamics and the quasibound $K^-pp$ cluster
}


\author{Akinobu Dot\'e         \and
        Tetsuo Hyodo            \\
        \and Wolfram Weise
}

\authorrunning{Akinobu. Dot\'e \and Tetsuo Hyodo \and Wolfram Weise} 

\institute{A. Dot\'e \at
              High Energy Accelerator Research Organization (IPNS/KEK), \\
 1-1 Oho, Tsukuba, Ibaraki, Japan, 3050801 \\
              \email{dote@post.kek.jp}           
           \and
           T. Hyodo \at
Physik-Department, Technische Universit\"at M\"unchen, 
D-85747 Garching, Germany; \\
Yukawa Institute for Theoretical Physics,
Kyoto University, Kyoto 606--8502, Japan\\
           \and
           W. Weise \at
Physik-Department, Technische Universit\"at M\"unchen, 
D-85747 Garching, Germany \\
}

\date{Received: date / Accepted: date}

\maketitle

\begin{abstract}
The prototype of a $\bar{K}$ nuclear cluster, $K^-pp$, has been investigated
using effective $\bar{K}N$ potentials based on chiral SU(3) dynamics. 
Variational calculation shows a bound state solution with shallow binding 
energy $B(K^-pp)=20\pm3$ MeV and broad mesonic decay width 
$\Gamma(\bar{K}NN \rightarrow \pi Y N)=40$ - $70$ MeV. The $\bar{K}N(I=0)$ 
pair in the $K^-pp$ system exhibits a similar structure as the 
$\Lambda(1405)$. We have also estimated the dispersive correction, $p$-wave 
$\bar{K}N$ interaction, and two-nucleon absorption width. 

\end{abstract}

\section{Introduction}

$\bar{K}$ nuclei (nuclear systems with a bound anti-kaon) have recently 
become a hot topic in hadron and nuclear physics. With a phenomenological 
$\bar{K}N$ potential, it was suggested that the $\bar{K}$ nuclei could exist 
as deeply bound states with small width~\cite{AY_2002}. Experiments performed
in search for such states have so far been inconclusive~\cite{Exps}. An 
important prototype is the $K^-pp$ system, the simplest $\bar{K}$-nuclear 
cluster. Recently this system has been studied using 
Faddeev~\cite{Faddeev:Shevchenko,Faddeev:Ikeda} and 
variational~\cite{ppK:Akaishi,DW-HYP06} approaches with $\bar{K}N$ 
interactions constrained by scattering data and properties of the 
$\Lambda(1405)$. 

An essential ingredient to study $\bar{K}$ nuclei is the $\bar{K}N$ 
interaction below threshold, which is only accessible through the 
subthreshold extrapolation of the amplitude adjusted to $\bar{K}N$ scattering
data. Theoretical guidance is required for this extrapolation. Here we report
on the study of a variational calculation of $K^-pp$ system~\cite{DHW:2008} 
using the effective $\bar{K}N$ interaction based on chiral SU(3) 
dynamics~\cite{Hyodo:2007jq}.

\section{Formalism}

The present variational investigation focuses on the $K^-pp$ system with spin
and parity $J^\pi=0^-$ and isospin $(T, T_z)=(1/2, 1/2)$, where the parity 
assignment includes the intrinsic parity of the antikaon. Our model wave 
function for this $K^-pp$ state, $|\Psi \rangle$, has two components:
\begin{eqnarray}
|\Psi \rangle & = &{\cal N}^{-1} [ \; |\Phi_+\rangle + C \, 
|\Phi_-\rangle \; ], \nonumber \\
& & |\Phi_+ \rangle \equiv  
\Phi_+ (\bm{r}_1, \bm{r}_2, \bm{r}_K) 
\; \left| S_N = 0 \right\rangle 
\times \; \left| \, \left[ \, [NN]_{T_N=1} \, \bar{K} \, \right]_{T=1/2, T_z=1/2} \right\rangle , \label{Phi+} \\
& & |\Phi_- \rangle \equiv  
\Phi_- (\bm{r}_1, \bm{r}_2, \bm{r}_K) 
\; \left| S_N = 0 \right\rangle 
\times \; \left| \, \left[ \, [NN]_{T_N=0} \, \bar{K} \, \right]_{T=1/2, T_z=1/2} \right\rangle , \label{Phi-}
\end{eqnarray}
where ${\cal N}^{-1}$ is a normalization factor. The first, second and third 
terms in Eqs. (\ref{Phi+}) and (\ref{Phi-}) correspond to the spatial wave 
function, the spin wave function of the two nucleons (assuming $S_N=0$), and 
the isospin wave function of the total system, respectively. We consider 
two different isospin states of the two nucleons ($T_N=1$ in 
$|\Phi_+ \rangle$ and $T_N=0$ in $|\Phi_- \rangle$), while the $\bar{K}NN$ 
system in both cases has total isospin and third component 
$(T, T_z)=(1/2, 1/2)$. The dominant contribution is the $T_N=1$ component
corresponding to the leading $K^-pp$ configuration. The mixing with the 
$T_N=0$ component is caused by the difference between the $\bar{K}N$ 
interactions in $I=0$ and $I=1$. The spatial part of the wave functions are 
products of single particle wave packets and two-particle correlation 
functions. The $NN$ correlation function permits an adequate treatment of a 
realistic $NN$ potential with its strong short-range repulsion. The 
parameters in the model wave function are determined by minimization of the 
energy.

The Hamiltonian used in the present study is of the form 
\begin{equation}
\hat{H} = \hat{T} + \hat{V}_{NN} + {\rm Re}\,\hat{V}_{\bar{K}N} - 
\hat{T}_{CM}~,  
\nonumber
\end{equation}
where $\hat{V}_{NN}$($\hat{V}_{\bar{K}N}$) stands for the $NN$($\bar{K}N$)
interaction. Here $\hat{T}$ is the total kinetic energy. The energy of the 
center-of-mass motion, $\hat{T}_{CM}$, is subtracted. As a realistic 
nucleon-nucleon interaction $\hat{V}_{NN}$ we choose the Argonne v18 
potential (Av18)~\cite{Av18}. We employ the central, $L^2$ and spin-spin 
parts of the Av18 potential for the singlet-even ($^1E$) and singlet-odd 
($^1O$) channel, since the total spin of the two nucleons is restricted to 
zero in our model. We use the $\bar{K}N$ interaction $\hat{V}_{\bar{K}N}$ 
derived from chiral SU(3) dynamics~\cite{Hyodo:2007jq}. This complex and 
energy-dependent interaction is parametrized by a Gaussian spatial 
distribution:
\begin{align}
    \hat{V}_{\bar{K}N} & = \hat{v}(\bar{K}N_1) +\hat{v}(\bar{K}N_2)
    \nonumber  \\
    \hat{v}(\bar{K}N) & =\sum_{I=0,1} \hat{P}_I (\bar{K}N) \times~ v_{\bar{K}N}^I (\sqrt{s}) \, \exp\left[-(\bm{r}_{\bar{K}}-\bm{r}_{N})^2 / a_s^2 \right], 
\nonumber
\end{align}
where $\hat{P}_I (\bar{K}N)$ is the isospin projection operator for the 
$\bar{K}N$ pair. The interaction strength $v_{\bar{K}N}^I (\sqrt{s})$ is a 
function of the center-of-mass energy variable $\sqrt{s}$ of the $\bar{K}N$ 
subsystem. The strength $v^{\bar{K}N,S}_I (\sqrt{s})$ and the range parameter
$a_s$ are systematically determined within the chiral coupled-channel 
approach. 

The energy dependence of the $\bar{K}N$ interaction requires the 
self-consistency in the variational procedure~\cite{DW-HYP06}. We introduce 
an auxiliary (non-observable) antikaon ``binding energy" $B_K$ to control the
energy $\sqrt{s}$ of the $\bar{K}N$ subsystem within the $K^-pp$ cluster. 
This $B_K$ is defined as 
\begin{equation}
- B_K \equiv \langle \Psi | \hat{H} | \Psi \rangle 
- \langle \Psi | \hat{H}_{N} | \Psi \rangle~,
\nonumber
\end{equation}
where $\hat{H}_{N}$ is the nucleonic part of the Hamiltonian. The relation 
between the $\bar{K}N$ two-body energy $\sqrt{s}$ and $B_K$ within the 
three-body system is not $a$ $priori$ fixed. In general, $\sqrt{s}$ can take 
values $M_N + m_K - \eta\, B_K$, where $\eta$ is a parameter describing the 
balance of the antikaon energy 
between the two nucleons of the $\bar{K}NN$ three-body system. One expects 
$1/2 \le\eta \le1$. The upper limit ($\eta = 1$) corresponds to the case in 
which the antikaon field collectively surrounds the two nucleons, a 
situation encountered in the limit of static (infinitely heavy) nucleon 
sources. In the lower limit ($\eta = 1/2$) the antikaon energy is split 
symmetrically half-and-half between the two nucleons. We investigate both 
cases and label them ``Type I" and ``Type II", respectively:
\begin{eqnarray}
    {\rm Type \; I \; :} ~~~~~~~~\sqrt{s}& = &M_N + m_K - B_K~~, 
    \nonumber\\
    {\rm Type \; II \; :} ~~~~~~~~\sqrt{s}& = &M_N + m_K - B_K/2~~. 
    \nonumber
\end{eqnarray}
Our calculation is then carried out such that self-consistency for $\sqrt{s}$
is achieved, namely, the $\sqrt{s}$ used in the effective $\bar{K}N$ 
potential is made to coincide with the $\sqrt{s}$ evaluated with the finally 
obtained wave function.

Due to the elimination of the $\pi\Sigma$ and $\pi\Lambda$ channels, the 
effective $\bar{K}N$ potential is complex. We perform the variational 
calculation with the real part of the potential to obtain the wave function. 
The decay width is then calculated perturbatively by taking the expectation 
value of the imaginary part of the potential: 
$\Gamma_M = -2 \; \langle \Psi | \, {\rm Im}\,\hat{V}_{\bar{K}N} \, | \Psi \rangle$, which represents the 
mesonic decay channels 
($\bar{K}NN \rightarrow \pi\Sigma N, \, \pi\Lambda N$). The dispersive effect
induced by the imaginary part of the potential and the non-mesonic absorption
width for $\bar{K}NN \rightarrow \Sigma N, \, \Lambda N$ are treated 
separately.

\section{Results}

\begin{figure}
  \includegraphics[width=0.50\textwidth]{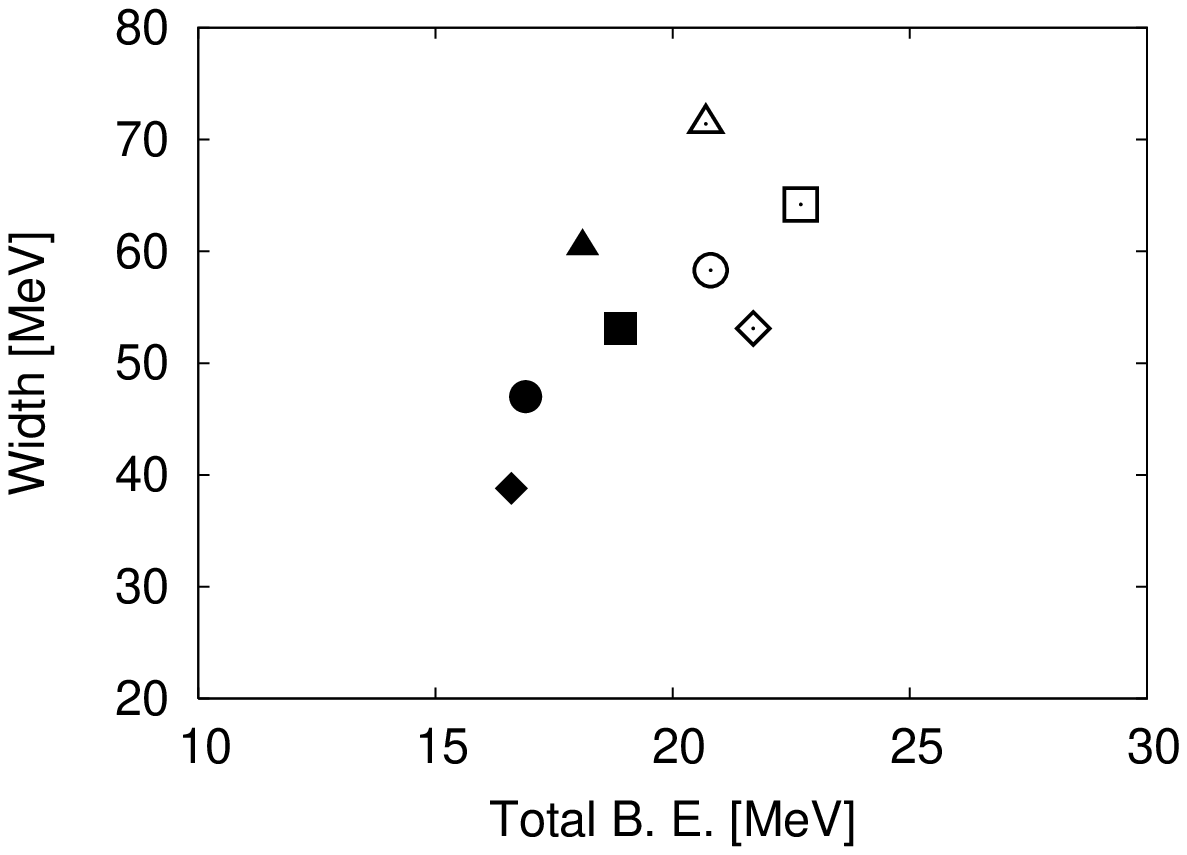}
  \includegraphics[width=0.45\textwidth]{KNIsoNormR2_v2.eps}
\caption{(Left panel) Distribution of total binding energy and mesonic decay 
width. The results of four models are shown with different symbols. Closed 
(Open) symbols indicate the Type I (Type II) ansatz. (Right panel) Normalized
and isospin-separated $\bar{K}N$ relative density in $K^-pp$ for a chiral 
model \cite{Hyodo:2002pk} with the Type I ansatz. Solid (dashed) line shows 
$I=0$ ($I=1$) $\bar{K}N$ density. Solid line with diamond shows the 
$\bar{K}N$ density of $\Lambda(1405)$ in the same model. All densities are 
displayed with $r^2$-multiplied. 
}
\label{fig:1}       
\end{figure}

\subsection{Structure of the $K^-pp$ system}

Here we present the results of the variational calculation. For an estimate 
of theoretical uncertainties, we have used four effective $\bar{K}N$ 
potentials derived from different versions of chiral models, and employed
both the Type I and the Type II ansatz, as described in Ref.~\cite{DHW:2008}.
In all cases the $K^-pp$ system turns out to be rather weakly bound as 
compared to previous calculations. As shown in the left panel of 
Fig.~\ref{fig:1}, the total binding energies range from 17 MeV to 23 MeV, and
the mesonic decay width $\Gamma_M$ ($\bar{K}NN \rightarrow \pi YN$) lies
between 40 and 70 MeV. The different versions of the chiral models give 
similar results within a relatively small window of uncertainties, while Type
II ansatz gives slightly deeper binding than Type I by a few MeV. The reason 
for the shallow binding is found in the relatively weak $\bar{K}N$ potentials
based on chiral dynamics. The chiral low energy theorem in the SU(3) 
meson-baryon sector dictates strong $\pi\Sigma$ attraction, and the 
coupled-channel dynamics locates the resonance structure in the $\bar{K}N$ 
amplitude at 1420 MeV, displaced from the 1405 MeV measured in the 
$\pi\Sigma$ spectrum. The binding energy of the isolated $\bar{K}N(I=0)$ 
system is about 12 MeV measured from $\bar{K}N$ threshold.  

Table \ref{tab:1} shows a typical result of $K^-pp$ calculated with a chiral 
model~\cite{Hyodo:2002pk} and the Type I ansatz. The results of the other 
cases under study are essentially the same. The mean distance between two 
nucleons, $R_{NN}$, is about 2.2 fm which is smaller than that of the 
deuteron (about 4 fm) and close to the $NN$ distance in normal nuclei, but 
the system is obviously not much compressed. 

It is interesting to compare the $\bar{K}N(I=0)$ component in $K^-pp$ with 
the $\Lambda(1405)$ as the $\bar{K}N(I=0)$ two-body quasibound state. The 
mean distance of the $\bar{K}N$ pair in $K^-pp$ is found to be close to that 
for $\Lambda(1405)$, namely $R^{\bar{K}N}_{I=0} \simeq 1.8$ fm and 
$R_{\bar{K}N} (\Lambda^*) \simeq 1.9$ fm. Calculating the expectation value 
of the relative $\bar{K}N$ orbital angular momentum, it turns out that the 
$\bar{K}N(I=0)$ pair is dominated by $s$-wave, just as the $\bar{K}N$ pair 
forming the $\Lambda(1405)$. The structure of the $\bar{K}N(I=0)$ pair in 
the $K^-pp$ system is thus similar to that of the $\Lambda(1405)$. The right 
panel in Fig. \ref{fig:1} shows the $\bar{K}N$ relative density distribution 
of $\bar{K}N(I=0$ and 1) components extracted from $K^-pp$ which are 
normalized to compare with that of $\Lambda(1405)$. Apparently, the 
distribution of $\bar{K}N(I=0)$ pair in $K^-pp$ is very similar to that of 
the $\bar{K}N$ two-body quasibound state.

\begin{table*}[tb]
\caption{
Detail of the result with a chiral model \cite{Hyodo:2002pk} and Type I 
ansatz. ``B. E. ($K^-pp$)'' and ``$\Gamma_{\rm M}$'' are total binding energy
and mesonic decay width of $K^-pp$. ``$R_{NN}$'' (``$R_{\bar{K}N}$'') is the 
$NN$ ($\bar{K}N$) relative distance. ``$R^{\bar{K}N}_{I=0 \; (1)}$'' is the 
$I=0 \; (1)$ $\bar{K}N$ relative distance. ``B. E. ($\Lambda^*$)'' and 
``$R_{\bar{K}N} (\Lambda^*)$'' are the binding energy and $\bar{K}N$ mean 
distance of isolated $I=0$ $\bar{K}N$ system. Energies and width are given 
in unit of MeV, while distances are given in fm. 
}
\label{tab:1}       
\begin{tabular}{cc|cc|cc||cc}
\hline \noalign{\smallskip} 
B. E. ($K^-pp$) & $\Gamma_{\rm M}$ & 
$R_{NN}$ &$R_{\bar{K}N}$ & $R^{\bar{K}N}_{I=0}$ & $R^{\bar{K}N}_{I=1}$ 
& B. E. ($\Lambda^*$) & $R_{\bar{K}N} (\Lambda^*) $  
\\
\noalign{\smallskip}
\hline \noalign{\smallskip}
16.9 & 47.0 & 
2.21 & 1.97 & 1.82 & 2.33  
& 11.5 & 1.86  \\
\noalign{\smallskip}
\hline \noalign{\smallskip}
\end{tabular}
\end{table*}

\subsection{Estimate of additional effects}

Based on the wavefunction obtained above, we estimate the following 
contributions to the results which have not been taken into account so far: 
1) dispersive corrections by the imaginary part of the potential, 2) effect 
of the $p$-wave $\bar{K}N$ interaction, and 3) decay width from the 
two-nucleon absorption process~\cite{DHW:2008}.

First, we consider the dispersive correction induced by the imaginary part of
the $\bar{K}N$ potential. This effect can be calculated explicitly for the 
two-body $\bar{K}N$ system, by comparing the bound state solution of the real
part of the potential with the resonance structure observed in the scattering
amplitude with original complex potential. Examining four chiral models, we 
find an attractive shift of the binding energy $6\pm 3$ MeV in the two-body 
$\bar{K}N$ system. We therefore estimate that the dispersive correction would
add another $\Delta B\lesssim 15$ MeV to the binding energy of the $K^-pp$ 
system.

Secondly, the contribution of the $p$-wave $\bar{K}N$ interaction is 
estimated perturbatively with the $p$-wave $\bar{K}N$ interaction:
\begin{eqnarray}
v^{p\rm{-wave}}_{\bar{K}N} (\bm{r}_{\bar{K}N}, \sqrt{s}) 
=
V^0_{\bar{K}N, p} (\sqrt{s}) \;
\,\nabla \exp[-\bm{r}_{\bar{K}N}^2/a_p^2]\, \nabla~.
\end{eqnarray}
The coefficient $V^0_{\bar{K}N, p} (\sqrt{s})$ is complex and a detailed 
expression is given in Ref.~\cite{DW-HYP06}. A prominent feature in the 
$p$-wave interaction is the $\Sigma(1385)$ resonance below the threshold.
Since the $K^-pp$ system is weakly bound and the energy variable $\sqrt{s}$ 
lies slightly above the $\Sigma(1385)$ resonance, the $p$-wave 
contribution to the binding energy is repulsive, about $-3$ MeV. The decay 
width is increased by 10 $\sim$ 35 MeV, because of the large imaginary part 
around the $\Sigma(1385)$ resonance structure. 

Next, we estimate the contribution of the two-nucleon absorption process
(non-mesonic decay width, $K^-pp \rightarrow YN$). The width is calculated 
with the correlated three-body density 
$\rho^{(3)}(\bm{r}_K,\bm{r}_1,\bm{r}_2)$ as
\begin{eqnarray}
    \Delta\Gamma_{abs}& (K^{-}pp\to YN) 
    = 
    \frac{2\pi  B_0}{\omega}
    \beta_{pp}(\omega) 
    \times \int d^3\bm{r}\int 
    d^3\bm{x}\,\,
    \rho^{(3)}(\bm{r},\bm{r},\bm{x})\, 
    G(\bm{x}-\bm{r};a) .
    \nonumber
\end{eqnarray}
This is a generalization of the formula for $K^-$ absorption on proton pairs
in a heavy nucleus, where the coupling constant is constrained by a global 
fit to the kaonic atom data~\cite{Mares:2006vk}. For the application to the 
few-body system, we modify the delta function type interaction to the finite 
range Gaussian form. This procedure is necessary to account for short range 
correlations of the nucleons in the few-body system, and physically motivated
by the underlying mechanism of the meson-exchange picture. Using the 
correlation density obtained from the wave function of the $K^-pp$, the two 
nucleon absorption width is estimated to be 4 - 12 MeV. 

\section{Summary and discussion}

We have investigated the $K^-pp$ system with a variational method, employing 
a realistic $NN$ potential (Av18 potential) and an effective $\bar{K}N$ 
potential based on chiral SU(3) dynamics. With theoretical uncertainties in 
the model, the binding energy and decay width of the $K^-pp$ turns out to be
\begin{equation}
{\rm B.E.} \, (K^-pp) = 20 \pm 3 \; {\rm MeV}, \quad 
\Gamma_M(K^-pp \rightarrow \pi Y N) = 40 \text{ - } 70 \; {\rm MeV}.
\nonumber
\end{equation}
As a consequence of the strong $\pi \Sigma$ interaction in chiral scheme, the
strength of the $\bar{K}N$ interaction is reduced and therefore we find a 
weakly bound state. The $\bar{K}N(I=0)$ pair in the obtained wave function 
of the $K^-pp$ system exhibits a similar structure as the $\bar{K}N$ two-body
quasibound state in vacuum. We have estimated corrections, such as dispersive
correction, the $p$-wave $\bar{K}N$ potential, and the two-nucleon absorption
process. Taking these effects into account, the total binding energy 
increases slightly and the total decay width becomes as large as 60 - 120 
MeV.

Our result should be compared with another three-body Faddeev calculation
with chiral interaction~\cite{Faddeev:Ikeda}, where the $K^-pp$ state was 
found with 80 MeV binding energy. While Faddeev approach treats the coupled 
channels explicitly, our variational calculation works by eliminating the 
$\pi\Sigma N$ channel. Although the two-body $\pi\Sigma$ dynamics is fully 
incorporated in the effective $\bar{K}N$ interaction, the dynamics of 
$\pi \Sigma N$ three-body system may generate additional attraction (see also
Ref.~\cite{IS:2008}). In the coupled-channel framework, the obtained state is
the mixture of the $\bar{K}NN$ and $\pi\Sigma N$ components, as the 
$\Lambda(1405)$ resonance in $\bar{K}N$-$\pi\Sigma$ system. In this sense, 
our strategy is to focus on the $\bar{K}NN$ component, and the present 
framework may not be sensitive to the $\pi\Sigma N$ component.

Based on a recent experimental analysis, a broad structure at about 100 MeV 
below the $\bar{K}NN$ threshold is reported~\cite{DISTO}, the maximum of 
which coincides with the $\pi\Sigma N$ threshold. In the chiral framework, 
such a broad state in the deep subthreshold 
 region would be interpreted in terms of $\pi\Sigma N$ dynamics, 
 driven by the strong $\pi\Sigma$ attraction. The present investigation
is however not capable to deal with the $\pi\Sigma N$ 
component, since we have eliminated this channel. While the new 
report~\cite{DISTO} is an interesting observation, more careful analysis is 
needed to answer the question about its detailed structure.

\begin{acknowledgements}
    This project is partially supported by BMBF, GSI, by the DFG excellence 
    cluster ``Origin and Structure of the Universe.", by the Japan Society 
    for the Promotion of Science (JSPS), and by the Grant for Scientific 
    Research (No.\ 19853500, 19740163) from the Ministry of Education, 
    Culture, Sports, Science and Technology (MEXT) of Japan. This research is
    part of the Yukawa International Program for Quark-Hadron Science. 
\end{acknowledgements}



\end{document}